\documentclass[prl,twocolumn,showpacs,nofootinbib]{revtex4}
\usepackage{epsf}
\usepackage{amsmath}

\def\beq{\begin{equation}}
\def\eeq{\end{equation}}
\def\bea{\begin{eqnarray}}
\def\eea{\end{eqnarray}}
\def\beqa{\begin{equation}\begin{array}{l}}
\def\eeqa{\end{array}\end{equation}}
\def\eqlab#1{\label{eq:#1}}
\def\figlab#1{\label{fig:#1}}

\def\eref#1{(\ref{eq:#1})}
\def\Eqref#1{Eq.~(\ref{eq:#1})}

\def\Figref#1{Fig.~\ref{fig:#1}}

\def\half{\mbox{\small{$\frac{1}{2}$}}}

\def\third{\mbox{\small{$\frac{1}{3}$}}}

\def\barr{\left(\begin{array}{c}}
\def\earr{\end{array}\right)}
\def\bmat{\left(\begin{array}{cc}}
\def\emat{\end{array}\right)}
\def\al{\alpha}
\def\be{\beta}
\def\ga{\gamma} 
\def\de{\delta} \def\De{\Delta}
\def\veps{\varepsilon}  

\def\la{\lambda} \def\La{{\Lambda}}

 \def\Si{{\it\Sigma}}
  \def\Th{\Theta}

\def\pa{\partial}

\def\pa{\partial}

\def\nn{\nonumber}

\def\lag{{\mathcal L}}

\def\mathscr{\mathcal}

\def\3d{3-D}

\def\ol#1{\overline{#1}}

\def\ceft{$\chi$EFT}

\begin{document}
\preprint{WM-05-117}
\preprint{JLAB-THY-05-393}

\title{Electromagnetic Nucleon-to-Delta Transition in Chiral Effective-Field Theory
}

\author{Vladimir Pascalutsa}
\email{vlad@jlab.org}
\author{Marc Vanderhaeghen}
\email{marcvdh@jlab.org}
\affiliation{Physics Department, The College of William \& Mary, Williamsburg, VA
23187, USA\\
Theory Group, Jefferson Lab, 12000 Jefferson Ave, Newport News, 
VA 23606, USA}

\date{\today}

\begin{abstract}
We perform a relativistic chiral effective-field theory calculation of
pion electroproduction off the nucleon ($e^-\, N \rightarrow e^-\, N\,\pi$)
in the $\Delta$(1232)-resonance region. 
After fixing the three low-energy constants, corresponding to the magnetic (M1), electric (E2),
and Coulomb (C2) $\gamma N\Delta $ couplings, our calculation provides a prediction
for the momentum-transfer  and pion-mass dependence of the $\gamma N\Delta $ form factors.
The prediction for the pion-mass dependence resolves the discrepancy
between the recent lattice QCD results and the experimental value for the ``C2/M1 ratio"
at low $Q^2$.   
\end{abstract}

\pacs{12.39.Fe, 13.40.Gp, 13.60.Le}%

\maketitle
\thispagestyle{empty}

The  $\De(1232)$-resonance, the first excited state of the nucleon,
dominates many nuclear phenomena at energies between the one- and two-pion
production thresholds. The electromagnetic excitation of the $\De$-resonance, 
the $\ga N \De$ transition, has recently received a lot of attention.
At low  momentum-transfer ($Q^2$) it highlights the role of the pion 
cloud~\cite{Mainz97,LEGS97,Bates01,Sato2, GrS96, DMT,PaT00}, 
whereas at larger $Q^2$ it probes the onset of the perturbative QCD regime~\cite{pQCD,Jlab}. 

The $\ga N \De$ transition is predominantly
of the magnetic dipole ($M1$) type which, in a simple quark-model picture, is described
by a spin flip of a quark in the $s$-wave state. Any $d$-wave admixture
in the nucleon {\it or} the $\Delta$ wave-functions allows for the electric- ($E2$) and Coulomb- ($C2$)
quadrupole transitions. Therefore by measuring these one is able to assess the presence
of the $d$-wave components and hence quantify to which extent the nucleon or the $\De$ wave-function
deviates from the spherical shape (``hadron deformation'')~\cite{deformation}.

The $\ga N\Delta$ transition has been 
accurately measured in the pion photo- and electro-production reactions 
\cite{Mainz97,LEGS97,Bates01,Jlab}. The $E2$ and $C2$
 are found to be relatively small, the ratios $R_{EM}=E2/M1$ and $R_{SM}=C2/M1 $ are at the level of a few percent.  
On the theoretical side, the most recent state-of-the-art lattice QCD study~\cite{Ale05} obtained a puzzling result:
the computed ratio $R_{SM}$  at low momentum-transfer appears to be significantly different from the observed value. 
It is important to note that the lattice calculations were done at larger pion masses, while the
result compared with experiment was obtained by a linear extrapolation to the physical pion mass. 

In this Letter we present a first chiral effective-field theory (\ceft)
calculation of pion photo- and electro-production on the nucleon in the
$\Delta$-resonance region. Besides finding a good agreement of our calculation with observables,
we are able to study the chiral behavior ($m_\pi$-dependence) of the $\ga N\De$ transition.
Our results show that {\em there is no} apparent discrepancy between the lattice data~\cite{Ale05}
and the experimental result for $R_{SM}$.

Our starting point is the relativistic chiral 
 Lagrangian of pion and nucleon
fields~\cite{GSS89} supplemented with the relativistic $\De$-isobar fields~\cite{PP03}.
 We organize the Lagrangian $\lag^{(i)}$, such that superscript $i$ stands for 
the power of electromagnetic 
coupling $e$ plus the number of derivatives
of pion and photon fields. Writing here only the relevant terms involving the spin-3/2 isospin-3/2
field $\psi^\mu$
of the $\De$-isobar we have (with antisymmetric products
of $\ga$-matrices:
$\ga^{\mu\nu}=\half[\ga^\mu,\ga^\nu]$,
$\ga^{\mu\nu\al}=
i\veps^{\mu\nu\al\be}\ga_\be\ga_5$):
\begin{subequations}
\eqlab{lagran}
\bea
\lag^{(1)}_\De &=&  \ol\psi_\mu \left(i\ga^{\mu\nu\al}\,D_\al - 
M_\De\,\ga^{\mu\nu}\right) \psi_\nu \nn\\
&+& \!\frac{i h_A}{2 f_\pi M_\De}\left\{
\ol N\, T_a \,\ga^{\mu\nu\la}\, (\pa_\mu \psi_\nu)\, D_\la \pi^a 
+ \mbox{H.c.}\right\} \\
\lag^{(2)}_\De &=&   \frac{3 i e g_M}{2M (M+M_\Delta)}\,\ol N\, T_3
\,\pa_{\mu}\psi_\nu \, \tilde F^{\mu\nu} \nn\\
&-& \!\frac{e h_A}{2 f_\pi M_\De}
\ol N\, T_a\,\ga^{\mu\nu\la} A_\mu \psi_\nu\, \pa_\la \pi^a + \mbox{H.c.},\\
\lag^{(3)}_\De &=&  \frac{-3 e}{2M (M+M_\Delta)}\, \ol N\, T_3\,
\ga_5 \left[ g_E\,\, (\pa_{\mu}\psi_\nu) F^{\mu\nu} \right. \nn\\
&+&\left.  \frac{g_C}{M_\De} \,
\ga^\al\,  (\pa_{\al}\psi_\nu-\pa_\nu\psi_\al)\,i\,\pa_\mu F^{\mu\nu}\right]+ \mbox{H.c.},
\eea
\end{subequations}
where $M\simeq 0.939$ and $M_\De\simeq 1.232$ GeV are, respectively, the nucleon and $\De$-isobar masses,
$N$ and $\pi^a\,\, (a=1,2,3)$ stand for the nucleon and pion fields, $D_\mu$ is the covariant 
derivative ensuring the electromagnetic gauge-invariance, $F^{\mu\nu}$ and $\tilde F^{\mu\nu}$
are the electromagnetic field strength and its dual,
$T_a$ are the isospin 1/2 to 3/2 transition matrices,  $f_\pi \simeq 92.4$ MeV is the pion decay constant. 
$\lag^{(1)}_\De$ contains the Rarita-Schwinger Lagrangian~\cite{RaS41} of a free spin-3/2 field 
formulated such that the number of spin degrees of freedom  is constrained
to the physical number.  The couplings in \Eqref{lagran}
are consistent with these constraints because of a spin-3/2 gauge symmetry~\cite{Pas98}. 

We next turn to  the power-counting for the pion electroproduction amplitude
using the ``$\de$-expansion'' scheme~\cite{PP03}. In this scheme 
the excitation energy of the $\De$-resonance: $\De\equiv M_\De-M\simeq 0.3$ GeV
is treated as a light scale,  so that for $\La\sim 1$ GeV representing the heavy scales in the theory,
we can use a small parameter $\de = \De/\La$. The other typical light scale of the theory,
the pion mass, is counted as two powers of the small parameter:  $m_\pi/\La\sim\de^2$.
The latter rule is the main distinction of this scheme from the previous power countings~\cite{SSE1,SSE2} which
count $\De$ and $m_\pi$ at the same order. This difference plays a crucial
role in separating the low-energy and resonance regimes, as well as in approaching
the chiral limit where $m_\pi$ vanishes while $\De$ remains finite.
Because of the distinction of $m_\pi$ and $\De$ the counting of a given diagram depends 
on whether the characteristic momentum $p$ is  
in the low-energy region ($p\sim m_\pi$) or in the resonance
region ($p\sim \De$). 
In the resonance region, one distinguishes the one-$\De$-reducible (O$\De$R) graphs~\cite{PP03}, see e.g.,
graph (a) in \Figref{diagrams}. Such graphs contain $\De$ propagators
which go as $1/(p-\De)$ and hence for $p\sim \De$ they are large and 
all need to be included. Their resummation amounts to 
dressing the $\De$ propagators so that they behave as $1/(p-\De-\Si)$. The self-energy 
$\Si$ begins at order $p^3$ and thus
a dressed O$\De$R propagator counts as $1/\de^3$.

\begin{figure}
\centerline{  \epsfxsize=8.8cm
  \epsffile{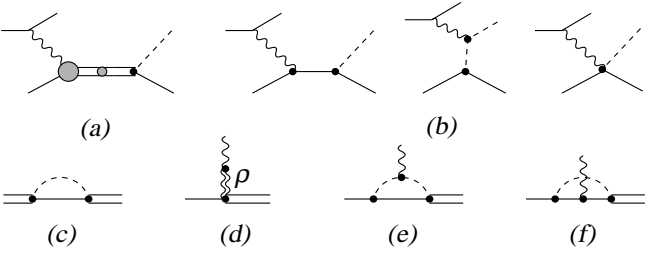} 
}
\caption{Diagrams for the $e N \to e \pi N $ reaction 
at NLO in the $\delta$-expansion, considered in this work. Double lines represent
the $\De$ propagators. The crossed nucleon-exchange graph is not shown in (b), but is included
in the calculation.}
\figlab{diagrams}
\end{figure}

The pion electroproduction amplitude to next-to-leading order (NLO) in the $\de$ expansion, in the
resonance region, is thus given by graphs in  \Figref{diagrams}(a) and (b), where the shaded
blobs in graph (a) include  corrections depicted in \Figref{diagrams}(c--f). The hadronic part of
graph (a) begins at ${\cal O}(\de^0)$ which here is the leading order. 
The Born graphs \Figref{diagrams}(b) contribute at ${\cal O}(\de)$.
We note that at NLO there are also vertex corrections of the type (e) and (f) with nucleon propagators in the loop
replaced by the $\De$-propagators. However, adopting the on-mass shell renormalizations and $Q^2\ll\La\De$, 
these graphs start to contribute at next-next-to-leading order (NNLO).

We have not shown the $\ga N\De$-vertex correction graph where the photon couples into the $\pi NN$ vertex,
because at this order the effect of this graph can fully be absorbed in the graphs  \Figref{diagrams}(e) and (f) 
by a field redifinition relating the pseudovector and pseudoscalar $\pi NN$ couplings. Having done that,
we compute graphs  \Figref{diagrams}(e) and (f) using the pseudoscalar coupling. 

The self-energy correction, \Figref{diagrams}(c), was computed previously~\cite{PV05}. 
In that calculation, the experimental value for the $\De$-resonance width fixes $h_A\simeq 2.85$.
To present the results for the vertex corrections we first consider the
general form of the $\gamma N \Delta$ vertex:
\begin{eqnarray}
 && \bar u_\alpha(p' ) \, \Gamma^{\alpha \mu}_{\ga N\De} \, u(p)   = 
\sqrt{\frac{3}{2}} \frac{ M_\Delta + M}{M \, [(M_\Delta + M)^2+Q^2]}  
\nonumber \\
&&\;\;\times  \; \bar u_\alpha(p^\prime) \, \left\{\, 
g_M(Q^2) \, \varepsilon^{\alpha \mu \kappa \lambda} \, 
p^{\prime}_\kappa \,  q_\lambda  \right. \nn\\
& & \;\; + \; g_E(Q^2) 
\left( q^\alpha \, p^{\prime \, \mu} -q \cdot p^\prime  \, g^{\alpha \mu} \right) i \gamma_5   \\
&&\;\;+ \; \left.  g_C(Q^2) 
\left( q^\alpha \, q^\mu - q^2 \, g^{\alpha \mu}  \right) i \gamma_5 
\right\} u(p) , \nn
\label{eq:diagndel1}
\end{eqnarray}
where $u_\al$ is the $\De$ vector-spinor, $u$ is the nucleon spinor, $q=p'-p$ is the
photon 4-momentum, $Q^2=-q^2$, and 
$g_M$, $g_E$, and $g_C$ are the form factors which at $Q^2=0$ 
are equal to the physical values of corresponding parameters  from Lagrangian \eref{lagran}. These
form factors relate to the conventional magnetic ($G_M^\ast$), electric ($G_E^\ast$) 
and Coulomb  ($G_C^\ast$) form factors of Jones and Scadron~\cite{Jones:1972ky} as follows:
\begin{eqnarray}
G_M^\ast &=& g_M \,+\frac{M_\De^2}{Q_+^2} \left(-\be_\ga \,g_E+  \bar Q^2 g_C\right), \nn\\ 
G_E^\ast &=& \frac{M_\De^2}{Q_+^2} \left(-\be_\ga \,g_E+ \bar Q^2  g_C\right), \\
G_C^\ast &=&-\frac{2 M_\De^2}{Q_+^2} \left(g_E +\be_\ga\,  g_C\right),\nn
\end{eqnarray}
where $Q_\pm=\sqrt{(M_\De\pm M)^2 +Q^2}$, $\bar Q^2 = Q^2/M_\De^2$,
$\be_\ga =\half  (1-r^2-\bar Q^2)$, with $r=M/M_\De$. The ratios $E2/M1$ and $C2/M1$ at the resonance
position can be expressed in terms of these form factors as:
\beq
\eqlab{ratios}
R_{EM}=-G_E^\ast/G_M^\ast\,,\;\;\;\; R_{SM}=-\mbox{$\frac{Q_+Q_-}{4M_\De^2}$}\, G_C^\ast/G_M^\ast .
\eeq

The one-loop corrections to the $\gamma N \Delta$ form factors are given by 
the graphs in \Figref{diagrams}(e) and (f). 
For example, the  ($\ol {MS}$-subtracted) result for the graph (e) in \Figref{diagrams} can be cast in the form:
\bea
g_M^{(e)} &=&  - C_{N\De} \! \int\limits_0^1 dy \, y \!\int\limits_0^{1-y} \!dx\,\ln {\cal M}^2 ,\nn \\
g_E^{(e)} &=&  - C_{N\De}  \! \int\limits_0^1 \!dy \, y \! \int\limits_0^{1-y} \!\!dx\,
\left\{ \ln {\cal M}^2 \right. \nn\\
&& \, \left. - 2 x \, [x \,r+(1-x-y)\,(1+r) ]\,  {\cal M}^{-2} \right\} ,\\
g_C^{(e)} &=&  -C_{N\De}  \! \int\limits_0^1 dy \, y \,(2 y-1)\!\!\int\limits_0^{1-y} \!\! dx\nn\\
&&\times\, [x r+(1-x-y)\,(1+r) ]\, {\cal M}^{-2} ,\nn
\eea
where 
$ {\cal M}^2 \equiv (x-\be )^2 -\la^2 +2\be_\ga  x y + \bar Q^2 y(1-y)- i\veps $, 
$\mu = m_\pi/M_\De$, $\be =\half  (1-r^2+\mu^2 )$, $\la^2=\be^2-\mu^2$,
 $C_{N\De}
 =4 g_A h_A Q_+^2 /[3(1+r)(8\pi f_\pi)^2]$, $g_A\simeq 1.26$. 
Analogous expressions are obtained for the graph \Figref{diagrams}(f). Alternatively, 
we have computed these graphs by using the sideways dispersion relations (see, e.g., \cite{HPV05}), 
and obtained identical results. 

The vector-meson diagram,  \Figref{diagrams}(d), contributes to NLO for $Q^2\sim \La\De$.We include
it effectively by giving the $g_M$-term a dipole $Q^2$-dependence (in analogy to how it is usually done
for the nucleon isovector form factor): $g_M\to g_M (1+Q^2/0.71\,\mbox{GeV}^2)^{-2}$. Analogous
effect for the $g_E$ and $g_C$ couplings begins at NNLO and is not included in the present calculation.

We now  present the electroproduction observables corresponding to 
the NLO amplitude of \Figref{diagrams}.
Denoting the invariant mass of the final $\pi N$ system by $s$, 
we restrict ourselves to the resonance kinematics:  $s = M_\Delta^2$.  
The $\gamma^* N \to \pi N$ cross section 
for unpolarized nucleons are expressed in terms of 5 response functions 
as~:
\begin{eqnarray}
\eqlab{xsecn}
\frac{d \sigma}{d \Omega_\pi} &=&
\frac{d \sigma_T}{d \Omega_\pi} 
+ \varepsilon \, \frac{d \sigma_L}{d \Omega_\pi} + \varepsilon \, \cos 2 \Phi \, 
\frac{d \sigma_{TT}}{d \Omega_\pi}\\
&& 
\hskip-1.3cm 
+\sqrt{2 \varepsilon  (1 + \varepsilon) }\, \cos \Phi \, 
\frac{d \sigma_{LT}}{d \Omega_\pi}  
+ h \sqrt{2 \varepsilon  (1 - \varepsilon) }\, \sin \Phi \, 
\frac{d \sigma_{LT}^ \prime}{d \Omega_\pi} , \nn
\end{eqnarray}
where $\Th_\pi$ and $\Phi$ are the pion polar and azimuthal c.m.\ angles, respectively, 
and $h$ denotes the electron helicity.

\begin{figure}[t]
\centerline{  \epsfxsize=8cm
  \epsffile{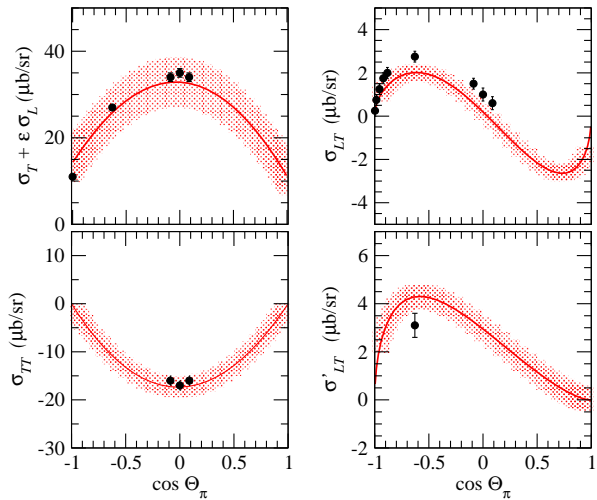} 
}
\caption{(Color online) \ceft\ NLO results for the  $\Th_\pi$ dependence of the 
$\ga^\ast p \to \pi^0 p$ cross sections at $\sqrt{s} = 1.232$~GeV and $Q^2$ = 0.127~GeV$^2$. 
The theoretical error bands are  described in the text. 
Data points are from BATES experiments~\cite{Bates01,Kunz:2003we}.
}
\label{crossections}
\end{figure}

In Fig.~\ref{crossections} we show our \ceft\ results for the different cross sections 
entering \Eqref{xsecn}. The only free parameters in this calculation are the low-energy constants
 from \Eqref{lagran}, which were chosen to yield the best description of the data as  
$g_M=2.88$, $g_E=-1.04$, $g_C = -2.36$. Within \ceft, we can estimate the theoretical uncertainty
of the NLO result  due to higher-order effects. 
The NNLO corrections to the amplitudes are expected to be of order of $\de^2$, $m_\pi/\La$, or $Q^2/\La^2$.
Therefore,  the theoretical uncertainty $R_{err}$ of an observable $R$, which involves a product of two amplitudes, 
is estimated as (taking here $\La=M$):
\beq
\eqlab{therror}
R_{err} = 2|R_{av}| \cdot \third \left( \delta^2 + \mbox{$\frac{m_\pi}{M}$}  + \mbox{$\frac{Q^2 }{ M^2}$} \right),
\eeq 
where $R_{av}$ is an average value of $R$. In  Fig.~\ref{crossections} the average is taken over the
range of  $\Theta_\pi$.
One sees that the NLO \ceft\ calculation, within its accuracy,
 is consistent with the experimental data for these observables. 

\begin{figure}
\centerline{  \epsfxsize=6.4cm
  \epsffile{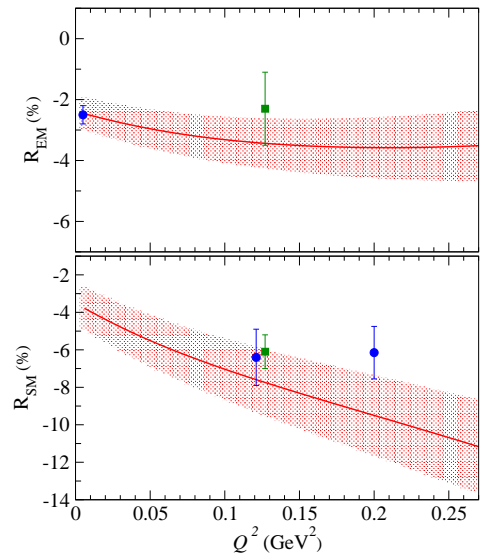} 
}
\caption{(Color online) $Q^2$ dependence of the NLO results for $R_{EM}$ (upper panel) and 
$R_{SM}$ (lower panel).
The blue circles are data points from MAMI for $R_{EM}$~\cite{Mainz97} , and 
 $R_{SM}$~\cite{Pospischil:2000ad,Elsner:2005cz}.  
The green squares are data points from BATES~\cite{Bates01}.}
\label{ratiosQ2}
\end{figure}

In Fig.~\ref{ratiosQ2} we show the $Q^2$ dependence of the ratios $R_{EM}$ and $R_{SM}$. Having 
fixed the low energy constants $g_M$, $g_E$ and $g_C$, the $Q^2$ dependence follows as a prediction.  
The theoretical uncertainty here (shown by  
the error bands) is estimated according to \Eqref{therror} with the average $R_{av}$
taken over the range  of $Q^2$ from 0 to 0.2~GeV$^2$. 
From the figure one sees that the NLO calculations are consistent with the experimental data for both
of the ratios.

\begin{figure}
\centerline{  \epsfxsize=6.5cm
  \epsffile{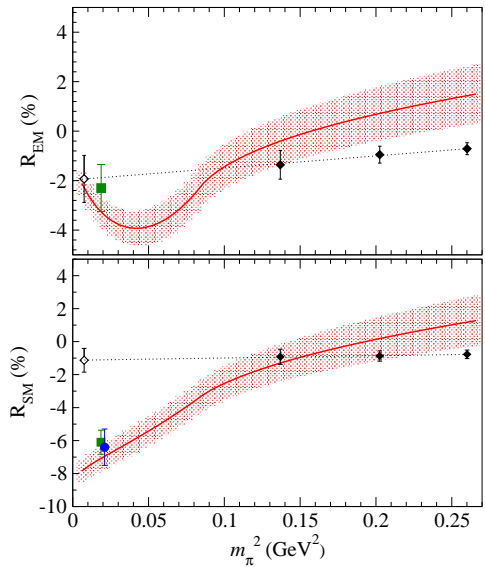} 
}
\caption{(Color online) $m_\pi$ dependence of the NLO results at $Q^2=0.1$ GeV$^2$ for 
 $R_{EM}$ (upper panel) and 
$R_{SM}$ (lower panel).
The blue circle is a data point from MAMI~\cite{Pospischil:2000ad}, 
the green squares are data points from BATES~\cite{Bates01}. 
The solid black diamonds 
are lattice calculations~\cite{Ale05}, 
whereas the dashed lines and open diamonds represent their extrapolation assuming linear dependence in $m_\pi^2$. }
\label{ratios}
\end{figure}

In Fig.~\ref{ratios} we show the $m_\pi$ dependence of the $\gamma N \De$ transition ratios, 
with the theoretical uncertainty  estimated according to \Eqref{therror} where $R_{av}$ is
taken over the range  of $m_\pi^2 $ from 0 to 0.15~GeV$^2$. 
The study of the $m_\pi$ dependence is crucial to connect to the  lattice QCD results, 
which at present  can only be obtained for larger pion masses. The recent
state-of-the-art lattice calculations of these ratios~\cite{Ale05}
use a {\it linear}, in the quark mass ($m_q\propto m_\pi^2$), {\it extrapolation}
to the physical point,  thus assuming that the non-analytic $m_q$-dependencies 
are  negligible. The thus obtained value for $R_{SM}$ at the physical $m_\pi$ value displays a large 
discrepancy with the  experimental result, as seen in Fig.~\ref{ratios}. 
However, our calculation demonstrates that the non-analytic dependencies are {\it not}
negligible. While
at larger values of $m_\pi$, where the $\Delta$ is stable, the ratios display a smooth 
$m_\pi$ dependence, at $m_\pi =\De $ there is an inflection point, and 
for  $m_\pi \leq \Delta$ the non-analytic effects are crucial, as was also observed for the $\De$-resonance
magnetic moment~\cite{Cloet03,PV05}.
The $m_\pi$ dependence obtained in \ceft\  clearly shows that
the lattice results for $R_{SM}$ may in fact be consistent with experiment.

In conclusion, we have performed a manifestly gauge- and Lorentz-invariant \ceft\
calculation of the $e N \to e  N \pi $ reaction
in the $\Delta(1232)$ resonance region. 
To NLO
in the $\delta$-expansion, the only free parameters entering the
calculation are the $\ga N\De$ couplings $g_M$, $g_E$, $g_C$ characterizing
the $M1$, $E2$, and $C2$ transitions. Our results agree well with 
recent high-precision data from MAMI and MIT-Bates at low $Q^2$.  
The \ceft\ framework plays a dual role, in that it allows for
an extraction of resonance parameters from observables {\em and} 
predicts their  $m_\pi$ dependence. In this way it may provide
 a crucial connection of present lattice QCD results obtained at unphysical values of $m_\pi$
to the experiment.
We have found that the opening of the $\De\to \pi N$ decay channel at $m_\pi = M_\De-M$
induces a pronounced  non-analytic
behavior of the $R_{EM}$ and $R_{SM}$ ratios. 
While the linearly-extrapolated lattice QCD results 
for $R_{SM}$ are in disagreement with experimental data, the \ceft\ prediction of
the non-analytic dependencies has allowed us to reconcile these results with experiment.
As the  next-generation lattice calculations of these quantities are on the way~\cite{Alexandrou:2005em}, 
the \ceft\ framework
presented here will, hopefully, complement these efforts.

\begin{acknowledgments}
This work is supported in part by DOE grant no.\
DE-FG02-04ER41302 and contract DE-AC05-84ER-40150 under
which SURA operates the Jefferson Laboratory.  
\end{acknowledgments}


\begin{thebibliography}{99}



\bibitem{Mainz97} R.~Beck {\em et al.},  Phys.\ Rev.\ Lett.\  {\bf 78}, 606 (1997); 
Phys.\ Rev.\ C  {\bf 61}, 035204 (2000).

\bibitem{LEGS97} G.~Blanpied  {\em et al.},  Phys.\ Rev.\ Lett.\  {\bf 79}, 4337 (1997).


\bibitem{Bates01}  C.~Mertz {\it et al.},
  Phys.\ Rev.\ Lett.\  {\bf 86}, 2963 (2001);
N.~F.~Sparveris {\it et al.}, {\it ibid.} {\bf 94}, 022003 (2005).

\bibitem{Sato2} S.~Nozawa, B.~Blankleider and T.-S.~H.~Lee,
Nucl.\ Phys.\ A {\bf 513}, 459 (1990);
 T. Sato and T. -S. H. Lee, Phys.\ Rev.\ C {\bf 54},
2660 (1996); {\it ibid.} {\bf 63}, 055201 (2001).

\bibitem{GrS96}
Y.~Surya and F.~Gross,
Phys.\ Rev.\ C {\bf 53}, 2422 (1996).


\bibitem{DMT} S.~Kamalov and S.~N. Yang, Phys. Rev. Lett. {\bf 83}, 4494 (1999);
S.~Kamalov {\it et al.},
Phys.\ Lett.\ B {\bf 522}, 27 (2001).

\bibitem{PaT00} V.~Pascalutsa and J.~A.~Tjon,
  Phys.\ Rev.\ C {\bf 70}, 035209 (2004);
 G.~Caia {\it et al.}, 
{\em ibid.} {\bf 70}, 032201(R) (2004).


\bibitem{pQCD}   C.~E.~Carlson,
  Phys.\ Rev.\ D {\bf 34}, 2704 (1986);
C.~Carlson and N.~Mukhopadhyay,
  Phys.\ Rev.\ Lett.\  {\bf 81}, 2646 (1998).

\bibitem{Jlab} V. V. Frolov {\it et al.}, Phys.\ Rev.\ Lett. {\bf
82}, 45 (1999);
K.~Joo {\it et al.}, {\em ibid.} {\bf 88}, 122001 (2002).


\bibitem{deformation}
N.~ Isgur, G.~Karl and R.~Koniuk, Phys.\ Rev.\ D {\bf 25}, 2394 (1982);
S.~Capstick and G.~Karl,  {\em ibid.} {\bf 41}, 2767 (1990);
G.~A.~Miller, Phys.\ Rev.\ C  {\bf 68}, 022201(R) (2003);
  A.~M.~Bernstein,
  Eur.\ Phys.\ J.\ A {\bf 17}, 349 (2003).

\bibitem{Ale05} C.~Alexandrou {\em et al.}, Phys.\ Rev.\ Lett. {\bf 94}, 021601 (2005).

\bibitem{GSS89} 
J.~Gasser, M.~E.~Sainio and A.~Svarc,
Nucl.\ Phys.\ B {\bf 307}, 779 (1988);
J.~Gegelia {\it et al.}, 
{J.~Phys.} G {\bf 29}, 2303 (2003).


\bibitem{PP03}
V.~Pascalutsa and D.~R.~Phillips,
 Phys.\ Rev.\ C {\bf 67}, 055202 (2003); {\it ibid.} {\bf 68}, 055205 (2003).



\bibitem{RaS41}
W.~Rarita and J.~S.~Schwinger,
Phys.\ Rev.\  {\bf 60}, 61 (1941).

\bibitem{Pas98}
V.~Pascalutsa,
Phys.\ Rev.\ D {\bf 58}, 096002 (1998);
Phys.\ Lett.\ B {\bf 503}, 85 (2001).
V.~Pascalutsa and R.~Timmermans,
Phys.\ Rev.\ C {\bf 60}, 042201(R) (1999).


\bibitem{SSE1}
E.~Jenkins and A.~V.~Manohar,
Phys.\ Lett.\ B {\bf 255}, 558 (1991);
 {\it ibid.} {\bf 259}, 353 (1991).

\bibitem{SSE2}
T.~Hemmert, B.~R.~Holstein and J.~Kambor,
 {\it ibid.} {\bf 395}, 89 (1997); 
  G.~Gellas {\it et al.},
  Phys.\ Rev.\ D {\bf 60}, 054022 (1999).


\bibitem{PV05}
V.~Pascalutsa and M.~Vanderhaeghen, 
  Phys.\ Rev.\ Lett.\  {\bf 94}, 102003 (2005).

\bibitem{Jones:1972ky}
H.~F.~Jones and M.~D.~Scadron,
Ann.~Phys.\  {\bf 81}, 1 (1973).


\bibitem{HPV05}
B.~R.~Holstein, V.~Pascalutsa, and M.~Vanderhaeghen,
  arXiv:hep-ph/0507016.

\bibitem{Kunz:2003we}
  C.~Kunz {\it et al.},
  Phys.\ Lett.\ B {\bf 564}, 21 (2003).

\bibitem{Pospischil:2000ad}
  T.~Pospischil {\it et al.},
  Phys.\ Rev.\ Lett.\  {\bf 86}, 2959 (2001).

\bibitem{Elsner:2005cz}
  D.~Elsner {\it et al.},
  arXiv:nucl-ex/0507014.

\bibitem{Cloet03}
R.~D.~Young, D.~B.~Leinweber and A.~W.~Thomas,
  Nucl.\ Phys.\ Proc.\ Suppl.\  {\bf 129}, 290 (2004).

\bibitem{Alexandrou:2005em}
  C.~Alexandrou {\it et al.},
  ``A study of the N to Delta transition form factors in full QCD,''
  arXiv:hep-lat/0509140.


\end{thebibliography}
\end{document}